\documentclass[journal]{IEEEtran}
\usepackage{graphicx}      
\usepackage[FIGTOPCAP]{subfigure}
\usepackage{natbib}        
\usepackage{enumitem}   
\usepackage{multirow}
\usepackage{theorem}
\usepackage[final]{listofsymbols}
\usepackage{amsmath}
\usepackage{amsfonts}
\usepackage{algorithm}
\usepackage[overload]{empheq}
\usepackage{hyperref}
\usepackage{xcolor}

\usepackage[acronym]{glossaries}
\newacronym{ocd}{\textsc{ocd}}{Optimality Condition Decomposition}
\newacronym{app}{\textsc{app}}{Auxiliary Problem Principle}
\newacronym{admm}{\textsc{admm}}{Alternating direction method of multipliers}
\newacronym{aladin}{\textsc{aladin}}{Augmented Lagrangian based Alternating Direction Inexact Newton method}
\newacronym{rapidpf}{rapid\textsc{pf}}{rapid prototyping for distributed power flow}
\newacronym{rapidpf+}{rapid\textsc{pf}+}{rapid prototyping for distributed power flow}
\newacronym{bfgs}{\textsc{bfgs}}{Broyden-Fletcher-Goldfarb-Shanno}
\newacronym{tsos}{\textsc{tso}s}{transmission system operators}
\newacronym{dsos}{\textsc{dso}s}{distribution system operators}
\newacronym{qp}{\textsc{qp}}{Quadratic Programing}
\newacronym{qcqp}{\textsc{qcqp}}{Quadratically Constrained Quadratic Program} 
\newacronym{licq}{\textsc{licq}}{linear independence constraint qualification} 
\newacronym{sosc}{\textsc{sosc}}{second order sufficient condition} 
\newacronym{scc}{\textsc{scc}}{strict complementarity conditions} 
\newacronym{kkt}{\textsc{kkt}}{Karush–Kuhn–Tucker} 
\newacronym{pf}{\textsc{pf}}{power flow}
\newacronym{opf}{\textsc{opf}}{optimal power flow}

\newacronym{nlp}{\textsc{nlp}}{Nonlinear Programming}

\usepackage{xspace}
\newcommand{\matlab}{\textsc{matlab}\xspace}

\newcommand{\norm}[1]{\left\lVert#1\right\rVert}
\newcommand{\n}{n}
\newcommand{\matpower}{\textsc{matpower}\xspace}
\newcommand{\runpf}{\textsc{runpf}\xspace}
\newcommand{\ipopt}{\textsc{ipopt}\xspace}

\newcommand{\aladinalpha}{\textsc{aladin}-$\alpha$\xspace}
\newcommand{\casadi}{\textsc{c}as\textsc{ad}i\xspace}
\newcommand{\fmincon}{\texttt{fmincon}\xspace}

\newcommand{\slack}{\textsc{ref}\xspace}
\newcommand{\pv}{\textsc{pv}\xspace}
\newcommand{\pq}{\textsc{pq}\xspace}
\newcommand{\ac}{\textsc{ac}\xspace}

\newtheorem{fact}{Fact}

\newtheorem{prop}{Proposition}

\newtheorem{theorem}{Theorem}

\begin{document}

\opensymdef
\newsym[rhomber of regions]{nregions}{\n^{\text{reg}}}
\newsym[Set of regions]{regions}{\mathcal{R}}
\newsym[Number of connections]{nconnections}{\n^{\text{conn}}}
\newsym[Number of core nodes]{ncore}{\n^\text{core}}
\newsym[Number of core nodes]{ncopy}{\n^\text{copy}}
\newsym[Number of power flow equations]{npf}{\n^\text{pf}}
\newsym[Number of buses]{nbus}{\n^\text{bus}}

\newsym[Voltage phase angle]{vang}{\textrm{$\theta$}}
\newsym[Voltage magnitude]{vmag}{v}
\newsym[Active power]{p}{p}
\newsym[Reactive power]{q}{q}

\newsym[State]{state}{x}

\newsym[Set of region]{setRegion}{\mathcal{R}}
\newsym[Set of bus]{setBus}{\mathcal{N}}

\newsym[Power flow problem]{pfproblem}{g}
\newsym[Power flow equations]{pfe}{g^{\text{pf}}}
\newsym[Bus specifications]{busspecs}{g^{\text{bus}}}

\newsym[]{gradient}{g}
\newsym[]{Hessian}{H}
\newsym[]{Jacobian}{J}

\newsym[]{DistOptLocalState}{x}
\newsym[]{DistOptUpdatedState}{z}
\newsym[]{SlackVState}{s}

\closesymdef

\title{\LARGE \bf Rapid Scalable Distributed Power Flow with Open-Source Implementation}
\author{Xinliang~Dai, Yichen~Cai, Yuning~Jiang, Veit Hagenmeyer\thanks{The authors acknowledge funding from the German Federal Ministry of Education and Research within the project \emph{MOReNet -- Modellierung, Optimierung und Regelung von Netzwerken heterogener Energiesysteme mit volatiler erneuerbarer Energieerzeugung}.}}


\maketitle

\begin{abstract}
This paper introduces a new method for solving the distributed \ac \acrfull{pf} problem by further exploiting the problem formulation. We propose a new variant of the \acrshort{aladin} algorithm devised
specifically for this type of problem. This new variant is characterized by using a reduced modelling method of the distributed  \ac \acrshort{pf} problem, which is reformulated as a zero-residual least-squares problem with
consensus constraints. This \acrshort{pf} is then solved by a Gauss-Newton based inexact \acrshort{aladin} algorithm presented in the paper. An open-source implementation of this algorithm, called \acrshort{rapidpf+}, is provided. Simulation results, for which the power system’s dimension varies from 53
to 10224 buses, show great potential of this combination in the aspects of both the computing
time and scalability.
\end{abstract}


\IEEEpeerreviewmaketitle

\section{Introduction}

The ongoing implementation of the energy transition leads to heterogeneous energy networks with numerous energy producers, energy consumers, transport, conversion and storage systems. Due to strongly varying renewable-based energy feed-ins and demands of the power system, new challenges arise in the aspect of power flow analysis, including \acrfull{pf} problems and \acrfull{opf} problems.

The conventional \acrshort{pf} problem is modeled as a system of nonlinear equations. Usually, it is solved by centralized methods, i.e., Gauss-Seidel or Newton-Raphson~\citep{grainger1999power}. However, the centralized approach requires one central entity, where all generation information and network topology data are collected. Sharing such data is unsatisfactory for system operators. In contrast to the centralized approach, the distributed approach first solves each decoupled sub-problem in its own local agent respectively, and then deals with a coupled problem in a central coordinator, in which only little information is acquired. As a result, the distributed approach not only preserves the information privacy and decision independence, but also decreases the vulnerability due to single-point-of failure~\citep{muhlpfordt2021distributed}.

The most well-known distributed algorithms for power flow analysis are \acrfull{ocd} proposed by~\cite{hug2009decentralized}, \acrfull{app} by~\cite{baldick1999fast}, and \acrfull{admm} by~\cite{erseghe2014distributed}. In this context, \acrshort{ocd} follows the idea of Lagrangian Relaxation Decomposition. Under certain assumptions, which cannot be guaranteed in general, it can converge to a solution with slight deviation to the optimizer. Different from \acrshort{ocd}, \acrshort{app} and \acrshort{admm} introduce Augmented Lagrangian Relaxation techniques to improve convergence behaviors, while \acrshort{admm} outperforms \acrshort{app} by reducing the communication effort. Although \acrshort{admm} has drawn significant attention for power flow analysis~\citep{erseghe2014distributed,kim2000comparison,guo2016case}, it normally takes quite a few iterates to approach a solution with moderate accuracy. Lately, \cite{sun2021two} proposed a two-level \acrshort{admm} for solving distributed \ac \acrshort{opf} problem with convergence guarantee. Nonetheless, the power flow model is formulated as a \acrfull{qcqp} problem at the expense of accuracy, and the algorithm converges to a modest accuracy slowly.

In addition, \cite{houska2016augmented} proposed the \acrfull{aladin} that is devised for non-convex problems with local convergence guarantee. It has found widespread application for power flow analysis of small- and medium-sized power systems~\citep{engelmann2018toward,meyer2019distributed,du2019distributed}. \acrshort{aladin} shares the same idea with \acrshort{admm}---update primal variables in an alternating fashion. However \acrshort{aladin} requires sensitivities information of sub-problems to build a second-order approximation in the coordinator. When using suitable Hessian approximation, \acrshort{aladin} can achieve locally quadratic convergence. In our previous work~\citep{muhlpfordt2021distributed}, an open-source \matlab code for \acrfull{rapidpf}\footnote{The code is available on https://github.com/KIT-IAI/rapidPF} is provided, in which the \ac \acrshort{pf} problem is reformulated as a zero-residual least-squares problem tailored for the \acrshort{aladin} to speed up the convergence---all the example cases can converge within half-dozen iterates. Nevertheless, the total computing time is not acceptable for large-scale problems due to the relative large dimension of the decoupled \acrshort{nlp} problem and the problematic code efficiency of \aladinalpha toolbox~\citep{engelmann2020aladin}.

The contribution of the present paper is two-fold. We propose a Gauss-Newton based \acrshort{aladin} algorithm for solving the zero-residual least-squares problem and a reduced modelling method for distributed \ac \acrshort{pf}. Based on them, we upgrade the open-source code of \acrshort{rapidpf}. The remainder of this paper is organized as follows: Section~\ref{sec:formulation} formulates the distributed \ac \acrshort{pf} as a zero-residual least-squares problem. Section~\ref{sec:alg} presents both the standard \acrshort{aladin} and the Gauss-Newton based \acrshort{aladin} algorithms. The upgrade of \acrshort{rapidpf}, called \acrshort{rapidpf+}, is described in~Section~\ref{sec::implementation}. The simulation results are compared and discussed in~Section~\ref{sec:results}.

\section{Problem Formulation}\label{sec:formulation}

This section introduces the distributed \ac \acrshort{pf} problem of polar voltage coordination and its zero-residual least-squares formulation. Before further discussion, we first introduce some nomenclature. For a power system, $\setRegion$ represents the set of regions, $\nregions$ is the number of regions and $\nconnections$ is the number of all the connecting tie lines between regions. In a specific region $\ell$, $\setBus_{\ell}$ is the set of all buses, whereas $\setBus^{core}_{\ell}$ and $\setBus^{copy}_{\ell}$ are the set of core and copy buses in this region $\ell$, respectively. 

\subsection{Distributed Power Flow}

The conventional \ac \acrshort{pf} problem seeks a deterministic solution to the steady-state operation of an \ac electrical power system by applying numerical analysis techniques~\citep{frank2016introduction}. Each bus in the system has four variables, i.e., voltage angle $\vang$, voltage magnitude $\vmag$, active power injection $\p$, and reactive power injection $\q$.
\begin{figure}[htbp!]
    \begin{center}
    \subfigure[Coupled system]{
        \includegraphics[width=0.2\textwidth]{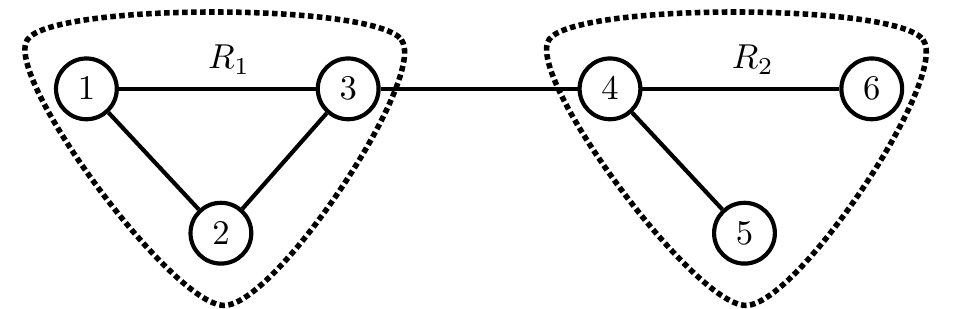}
        \label{fig:coupled system}
    }\\
    \subfigure[Decoupled region 2]{
        \includegraphics[width=0.15\textwidth]{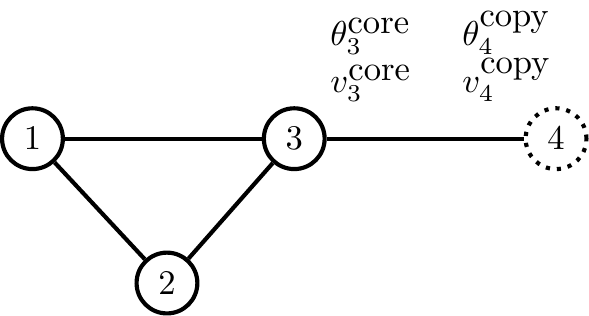}
        \label{fig:region1}
    }
    \hskip 20pt
    \subfigure[Decoupled region 2]{
        \includegraphics[width=0.15\textwidth]{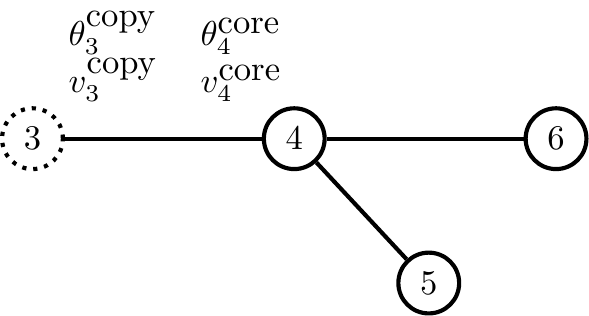}
        \label{fig:region2}
    }
    \end{center}
    \caption[Optional caption for list of figures]{Decomposition by sharing components for a two-region system}\label{fig:example}
\end{figure}

\begin{fact}\label{fact::solutions}
Genetically, there are multiple mathematically valid solutions to a power flow problem, but only one solution has physical meaning~\citep{frank2016introduction}.This results, e.g., from the periodic voltage angle $\vang$, and the respective trigonometric functions. 
\end{fact}          

In order to apply a distributed algorithm, reformulation of the \ac \acrshort{pf} problem is necessary. In terms of partitioning the power system, we share the components between neighboring regions to ensure physical consistency. As an example, we take the 6-bus system with 2 regions, shown in \autoref{fig:example}. The coupled system, shown in \autoref{fig:example}(a), has been partitioned into 2 local regions. To solve the \ac \acrshort{pf} problem in region $R_1$, besides its own buses \{1,2,3\} called \textit{core buses}, the complex voltage of bus \{4\} from neighboring region $R_2$ is required. Hence, for the sub-problem of region $R_1$, we create an auxiliary bus \{4\} called \textit{copy bus}, along with its own \textit{core bus}, to formulate a self-contained \ac \acrshort{pf} problem.

Then, affine consensus constraints of the connecting tie line are added to ensure consistency of the copy bus with its original core bus in the neighboring region. The consensus constraints of the example case in~\autoref{fig:example} can be written as
\begin{subequations}
    \begin{align}
        \vang^{\textrm{core}}_3 = \vang^{\textrm{copy}}_3,&\;\vang^{\textrm{core}}_4 = \vang^{\textrm{copy}}_4\\
        \vmag^{\textrm{core}}_3 = \vmag^{\textrm{copy}}_3,&\;\vmag^{\textrm{core}}_4 = \vmag^{\textrm{copy}}_4
    \end{align} 
\end{subequations}

In a specific region $\ell$, the power flow equations be represented as
\begin{subequations}\label{eq:power flow equation}
    \begin{align}
    \p_i^g-\p_i^l &=\vmag_i \sum_{k\in\setBus_{\ell}} \vmag_k \left( G_{ik} \cos\vang_{ik} + B_{ik} \sin\vang_{ik} \right)\\
    \q_i^g-\q_i^l &=\vmag_i \sum_{k\in\setBus_{\ell}} \vmag_k \left( G_{ik} \sin\vang_{ik} - B_{ij} \cos\vang_{ik} \right)
    \end{align}
\end{subequations}
for all \textit{core bus} $i\in \setBus^\textrm{core}_{\ell}$ with the angle difference between buses $\vang_{ik} = \vang_i-\vang_k$, complex generation $s^g = \p^g + \mathrm{j} \q^g$, complex load $s^l = \p^l + \mathrm{j} \q^l$, complex components of the bus admittance matrix entries $Y_{ik} = G_{ik} + \mathrm{j} B_{ik}$. These equations can also be written as residual function
\begin{equation}\label{eq:power flow residual}
    r_{\ell}(\chi_{\ell})=0
\end{equation}
where $r_{\ell}:\mathbb{R}^{2\ncore_\ell+2\ncopy_\ell} \rightarrow{} \mathbb{R}^{\npf_\ell}$ with its components $r_{\ell,m}$, i.e., the $m$-th power flow residual in region $\ell$. Note that the number of power flow equations $\npf_\ell = 2\ncore_\ell$ in all local region.

Hence, the distributed \ac \acrshort{pf} problem can be represented as a system of nonlinear equations and affinely coupled consensus equations as follows
\begin{subequations}~\label{eq:feasibility problem}
    \begin{align}
         r_{\ell}(\chi_{\ell})&=0,\;\forall\ell\in\setRegion\\
         \sum_{\ell\in\setRegion}A_\ell \chi_{\ell} &=A\chi=b. 
    \end{align}
\end{subequations}
with $\chi = (\chi_1^\top, \chi_2^\top,\cdots,\chi_{\nregions}^\top)^\top$

\subsection{Least-Squares Formulation}

Following~\cite{muhlpfordt2021distributed}, we reformulate the distributed \ac \acrshort{pf} problem~\eqref{eq:feasibility problem} in a standard least-squares formulation with affine consensus constraint
\begin{subequations}\label{eq::dist ls problem}
    \begin{align}
        \min_{\chi}\quad f(\chi):=&\sum_{\ell\in\mathcal{R}} f_\ell(\chi_{\ell}) = \frac{1}{2}\sum_{\ell\in\mathcal{R}}\norm{r_\ell(\chi_{\ell})}_2^2\\
        \textrm{s.t.}  \quad  A \chi =&\; b\;\;\mid\lambda\label{eq::ls::consensus}
    \end{align}
\end{subequations}
with the consensus matrix $A = (A_1, A_2, \cdots, A_{\nregions})$ and the state $\chi = (\chi_1^{\top},\chi_2^{\top},\cdots,\chi_{\nregions}^{\top})^{\top}$ .

\begin{prop}\label{prop:lambda=0}
Let the power flow problem~\eqref{eq:power flow equation} be feasible, i.e., a primal solution $\chi^{\ast}$ to the problem~\eqref{eq::dist ls problem} exists such that the power flow residual $r_\ell(\chi^{\ast}_\ell)=0$ for all $\ell\in\mathcal{R}$ bounded by consensus constraint~\eqref{eq::ls::consensus}, and let \acrfull{licq} holds at $\chi^{\ast}$. Then the dual variable $\lambda^{\ast} = 0$ with the primal solution $\chi^{\ast}$ satisfies the \acrshort{kkt} conditions, i.e., $(\chi^{\ast}, \lambda^{\ast} = 0)$ is a \acrshort{kkt} point.
 \end{prop} 

\subsection{Sensitivities}\label{sec::senstivities}

The derivatives of the objective $f_\ell(\chi_\ell)$ can be expressed as
\begin{subequations}
    \begin{align}
        \nabla f_\ell(\chi_{\ell}) &= J_{\ell}(\chi_{\ell})^{\top} r_\ell(\chi_{\ell})\\
        \nabla^2 f_\ell(\chi_{\ell}) &= J_{\ell}(\chi_{\ell})^{\top} J_{\ell}(\chi_{\ell}) +  Q_{\ell}(\chi_{\ell})
    \end{align}
\end{subequations}
with 
\begin{subequations}
    \begin{align}
    J_{\ell}(\chi_{\ell}) &= 
    \begin{bmatrix}
    \nabla r_{\ell,1},\nabla r_{\ell,2},\cdots,\nabla r_{\ell,\npf}
    \end{bmatrix}^{\top}\\
    Q_{\ell}(\chi_{\ell}) &= \sum_{m=1}^{\npf}r_{\ell,m}(\chi_{\ell}) \nabla^2 r_{\ell,m}(\chi_{\ell}).
    \end{align}
\end{subequations}

In practice, the first term $J_{\ell}(\chi_{\ell})^{\top} J_{\ell}(\chi_{\ell})$ of the second order derivative dominates the second one $Q_{\ell}(\chi_{\ell})$, either because the residuals $r_{\ell,m}$ are close to affine near the solution, i.e., $\nabla^2 r_{\ell,m}$ are relatively small, or because of small residuals ~\citep{nocedal2006numerical}. For solving zero-residual least-squares problem, we hence chose the so-called Gauss-Newton approximation
\begin{equation}
    \nabla^2 f_\ell(\chi_{\ell}) \approx J_{\ell}(\chi_{\ell})^{\top} J_{\ell}(\chi_{\ell})\label{eq:gauss-newton}
\end{equation}

\section{Algorithm}\label{sec:alg}

This section presents the standard \acrshort{aladin} algorithm and its new variant for zero-residual least-squares problems.

\subsection{Standard ALADIN}

\cite{houska2016augmented} introduced a novel algorithm, i.e., \acrshort{aladin}, to handle distributed nonlinear programming. \acrshort{aladin} for problem~\eqref{eq::dist ls problem} is outlined in~\autoref{alg:standard-aladin}. The algorithm has two main steps, i.e., a decoupled step~\ref{alg:decoupled} and a consensus step~\ref{alg:consensus}.  Pursuing the idea of augmented Lagrangian, the local problem is formulated as~\eqref{eq:aladin_nlp} in step~\ref{alg:decoupled}, where $\rho$ is the penalty parameter and $\Sigma_\ell$ is the positive definite scaling matrix for the region $\ell$. Based on the result from local \acrshort{nlp}s~\eqref{eq:aladin_nlp}, the \acrshort{aladin} algorithm terminates if both the primal and the dual residuals are smaller than tolerance $\epsilon$
\begin{equation}\label{eq:termination}
            \norm{\sum_{\ell\in\mathcal{R}} A_\ell \DistOptLocalState_\ell - b}_\infty \leq \epsilon\;\textrm{and}\;\max_{\ell}\norm{\Sigma_{\ell}(\DistOptLocalState_\ell -\DistOptUpdatedState_\ell )}_\infty\leq \epsilon
\end{equation}
\begin{algorithm}[htbp!]
    \caption{\acrshort{aladin}(standard)}\label{alg:standard-aladin}
    \small
    \textbf{Initialization: $\lambda,\rho,\mu$, $\DistOptUpdatedState_\ell$, $\Sigma_\ell\succ 0$ for all $\ell \in \mathcal{R}$, }\\
    \textbf{Repeat:}
    \begin{enumerate}[label=(\roman*)]
        \item Solve decoupled \acrshort{nlp}s
        \begin{align}\label{eq:aladin_nlp}
            \underset{\DistOptLocalState_\ell}{\operatorname{min}}\;\;f_\ell(\DistOptLocalState_\ell) + \lambda^{\top} A_\ell \DistOptLocalState_\ell + \frac{\rho}{2}\norm{\DistOptLocalState_\ell-\DistOptUpdatedState_\ell}^2_{\Sigma_\ell}
        \end{align}
        and compute local sensitivities for all $\ell \in \mathcal{R}$ ~\label{alg:decoupled}
        \begin{align}\label{eq:aladin_sens}
                \gradient_\ell = \nabla f_\ell^{\phantom{\ell}}(\DistOptLocalState_\ell)\;\textrm{and}\;\Hessian_\ell \approx \nabla^2 f_\ell^{\phantom{\ell}}(\DistOptLocalState_\ell)
        \end{align}
        
        \item Check termination condition~\eqref{eq:termination}\\
        \item Solve coupled \acrshort{qp} ~\label{alg:consensus}
            \begin{subequations}\label{eq:aladin_qp}
                \begin{align}
                    \underset{\Delta \DistOptLocalState, \SlackVState  }{\operatorname{min}}&\;\;\frac{1}{2} \Delta \DistOptLocalState^\top \Hessian \Delta \DistOptLocalState +\gradient^\top \Delta \DistOptLocalState +\lambda^{\top}\SlackVState + \frac{\mu}{2}\norm{s}_2^2\\                    
                    \text{s.t.}&\;\;A\left( \DistOptLocalState+\Delta\DistOptLocalState \right) = b + s
                \end{align}
            \end{subequations}
        where Hessian $H=\text{diag}\{H_\ell\}_{\ell\in\mathcal{R}}$ and gradient $g$ with components $g_\ell$\\
        \item Update primal and dual variables with full-step ~\label{alg:updating}
            \begin{subequations}
                \begin{align}
                    \DistOptUpdatedState^{+} &= \DistOptLocalState + \Delta \DistOptLocalState,\label{eq::updating::primal}\\
                    \lambda^{+} &= \lambda^{QP}.
                \end{align}
            \end{subequations}
    \end{enumerate}
\end{algorithm}
Compared with a simple averaging step of \acrshort{admm} in the coordinator, \acrshort{aladin} based on curvature information~\eqref{eq:aladin_sens} builds a coupled \acrshort{qp}~\eqref{eq:aladin_qp} to coordinate the results of the decoupled step from all regions. Additionally, a slack variable $s$ is introduced in the consensus step to ensure feasibility of the coupled~\acrshort{qp}. Consequently, \acrshort{aladin} achieves fast and guaranteed convergence. A detailed proof of local convergence can be found in~\cite{houska2016augmented}. 

\subsection{Gauss-Newton based inexact ALADIN}

Based on the framework of standard \acrshort{aladin}, we propose a tailored version specific for solving zeros-residual least-squares problem in the present paper, see~\autoref{alg:gn-aladin}. Since optimal values of Lagrangian multipliers are equal to zero $\lambda^{\ast}=0$ according to \textit{Proposition~\ref{prop:lambda=0}}, the Lagrangian terms in \eqref{eq:aladin_nlp}\eqref{eq:aladin_qp} can be neglected by fixing dual iterates $\lambda=0$ at the cost of convergence rate. In this way, both coupled and decoupled steps can be viewed as adding a residual to the original problems respectively, and can be solved by equivalent linear systems efficiently.

\begin{algorithm}[htbp!]
    \caption{inexact \acrshort{aladin}(Gauss-Newton)}\label{alg:gn-aladin}
    \small
    \textbf{Initialization: $\lambda,\rho,\mu$, $\DistOptUpdatedState_\ell$, $\Sigma_\ell\succ 0$ for all $\ell \in \mathcal{R}$, }\\
    \textbf{Repeat:}
    \begin{enumerate}[label=(\roman*)]
        \item Solve decoupled linear systems and update primal variables $\DistOptLocalState_\ell$
        \begin{equation}\label{eq::inexact::decoupled}
            \left(\Jacobian_\ell^{\DistOptUpdatedState\top} \Jacobian^{\DistOptUpdatedState}_\ell +  \rho I\right) p_\ell = - \Jacobian_\ell^{\DistOptUpdatedState\top} r_\ell^{\DistOptUpdatedState}
        \end{equation}
        with Gauss-Newton step $p_{\ell}^{} = x_{\ell}-z_{\ell}$, as well as compute local sensitivities for all $\ell \in \mathcal{R}$
        \begin{equation}
                \gradient_\ell = \Jacobian_{\ell}(\hat{\DistOptLocalState}_\ell)^\top r_{\ell}(\hat{\DistOptLocalState}_\ell)\;\textrm{and}\;\Hessian_\ell = \Jacobian_{\ell}(\hat{\DistOptLocalState}_{\ell})^{\top} \Jacobian_{\ell}(\hat{\DistOptLocalState}_{\ell})
        \end{equation}
        \item Check termination condition~\eqref{eq:termination}\\
        \item Solve the linear system of coupled \acrshort{qp}
            \begin{align}\label{eq::equivalent-QP}
                \left(\Hessian + \mu A^{\top} A\right) \Delta \DistOptLocalState = - \mu A^{\top} \left(A \hat{\DistOptLocalState}-b\right) - \gradient
            \end{align}
        where Hessian $H=\text{diag}\{H_\ell\}_{\ell\in\mathcal{R}}$ and gradient $g$ with components $g_\ell$\\
        \item Update primal variables with full step
        \begin{equation}
            \DistOptUpdatedState^{+} = \hat{\DistOptLocalState} + \Delta \DistOptLocalState.
        \end{equation}
    \end{enumerate} 
\end{algorithm}
For the decoupled step~\ref{alg:decoupled}, the objective function~\eqref{eq:aladin_nlp} can be approximated by a quadratic model by applying the Gauss-Newton method
\begin{equation}
    M_{\ell}(p_\ell) = \frac{1}{2}p_{\ell}^{\top} \left( J_\ell^{\DistOptUpdatedState \top}  J_\ell^{\DistOptUpdatedState}+ \rho I \right) p_{\ell} + J_\ell^{\DistOptUpdatedState \top} r_\ell^\DistOptUpdatedState p_{\ell} +f_\ell(z_\ell) 
\end{equation}
with Gauss-Newton step ${p}_\ell = x_\ell - z_\ell$, Jacobian matrix $\Jacobian_\ell^{\DistOptUpdatedState} = \Jacobian_\ell(\DistOptUpdatedState_\ell)$ and residual vector $r_\ell^{\DistOptUpdatedState} = r_\ell(\DistOptUpdatedState_\ell)$ at the initial point $\DistOptUpdatedState_\ell$ in every iterate. Accordingly, the decoupled \acrshort{nlp}~\eqref{eq:aladin_nlp} is solved by a linear system~\eqref{eq::inexact::decoupled}, where $x_{\ell}$ is an inexact solution to this problem.


For the coupled step~\ref{alg:consensus}, the objective function can be rewritten as
\begin{equation}\label{eq::inexact::qp}
    \underset{\Delta \DistOptLocalState}{\operatorname{min}}\;\;\frac{1}{2} \Delta \DistOptLocalState^\top \Hessian \Delta \DistOptLocalState +\gradient^\top \Delta \DistOptLocalState  + \frac{\mu}{2}\norm{A\left(\hat{\DistOptLocalState}+\Delta\DistOptLocalState \right) - b}_2^2
\end{equation}
In the corresponding linear system~\eqref{eq::equivalent-QP}, $\Delta x$ in coupled step~\ref{alg:consensus} is locally equivalent to a standard Gauss-Newton step of the original coupled problem~\eqref{eq::dist ls problem}, where the slack variable $s = A\left( \hat{\DistOptLocalState}+\Delta\DistOptLocalState \right) - b$ can be viewed as an additional weighted residual. 

In the present paper, we focus on the local convergence due to~\textit{Fact~\ref{fact::solutions}} and good initial guess provided by \matpower. The local convergence indicates that the starting point and the iterates are all located in a small neighborhood of the optimizer, within which the solution has physical meaning. The convex set $\Omega$ concludes all the points in the bounded neighborhood. Besides, the objective $f$ of the original coupled problem~\eqref{eq::dist ls problem} is second order continuously differentiable according to Section~\ref{sec::senstivities}, and $\norm{\nabla^2 f(x)}$ is bounded for all $x\in\Omega$. Then, there exists a constant $L>0$
\begin{equation}\label{eq::lipschitz}
    \norm{\nabla f(x)-\nabla f(z^{\ast})}=\norm{\nabla^2 f(\tilde{x})} \norm{x-z^{\ast}}\leq L\norm{x-z^{\ast}}
\end{equation}
with $\tilde{x} = x - t(x-z^{\ast})\in\Omega$ for some $t\in(0,1)$. Hence, the function $f$ is twice Lipschitz-continuously differentiable in the neighborhood $\Omega$. 

Before discussing further about the convergence property, we introduce a regularity and some nomenclature first: A \acrshort{kkt} point is called \textit{regular} if \acrfull{licq}, \acrfull{scc} and \acrfull{sosc} are satisfied. For the analysis of local decoupled step~\ref{alg:decoupled}, we introduce $\bar{x}$ as the exact solution and $\hat{x}$ as the inexact solution of the decoupled \acrshort{nlp}s~\eqref{eq:aladin_nlp}, whereas $x^{\ast}=z^{\ast}$ is the primal optimizer of the original coupled problem~\eqref{eq::dist ls problem}.

Next, let's turn to the local convergence property of \autoref{alg:gn-aladin}.

\begin{theorem}\label{thm::convergence}
Let the minimizer $(x^{\ast} = z^{\ast},\lambda^{\ast}=0)$ be a regular \acrshort{kkt} point of problem~\eqref{eq::dist ls problem}, let the initial guess located in the small neighborhood of the optimizer $\Omega$, and let $\mu$ sufficient large such that $\frac{1}{\mu}\leq O(\norm{\hat{x}-z^{\ast}})$, then the iterates $\hat{x}$ of \autoref{alg:gn-aladin} converge quadratically to a local solution.
\end{theorem}
Proof of \textit{Theorem~\ref{thm::convergence}} can be established by three steps, following the analysis in \textit{Appendix} by \cite{engelmann2018toward}. First, due to the fact that the local inexact solution $\hat{x}_\ell$ is obtained by Gauss-Newton method, the $\hat{x}$ is a linear contraction to the exact solution $\bar{x}$, i.e., there exists a constant $\eta_{1}>0$ such that
\begin{equation}\label{eq::convergence::inexat::nlp}
    \norm{\hat{x}-\bar{x}}\leq \eta_{1} \norm{z - \bar{x}}.
\end{equation}

Second, from \textit{Lemma~3} of \cite{houska2016augmented}, we have
\begin{equation}\label{eq::convergence::decoupled}
    \norm{\bar{x}-z^{\ast}}\leq \eta_{2}\norm{z-z^{\ast}},\;\exists \eta_{2}>0
\end{equation}
This differs from standard \acrshort{aladin} by a fixed dual variable $\lambda = 0$.

Third, because the coupled step of \autoref{alg:gn-aladin} is a standard Gauss-Newton step of the original coupled problem~\eqref{eq::dist ls problem}, as well as the Lipschitz continuity of $f$ and sufficient large $\mu$ such that $\frac{1}{\mu}\leq O(\norm{\hat{x}-z^{\ast}})$, we obtain the following inequality according to the convergence analysis of the standard Gauss-Newton method~\citep[Section 10.3]{nocedal2006numerical}
\begin{equation}
    \norm{\DistOptUpdatedState^{+}- \DistOptUpdatedState^{\ast}} \leq \norm{H(\DistOptUpdatedState^{\ast})^{-1}Q(\DistOptUpdatedState^{\ast})}\norm{\DistOptLocalState - \DistOptUpdatedState^{\ast}} + O(\norm{\DistOptLocalState - \DistOptUpdatedState^{\ast}}^2)
\end{equation}
with $Q=\text{diag}\{Q_\ell\}_{\ell\in\mathcal{R}}$. For problem~\eqref{eq::dist ls problem}, all the optimal residuals are equal to zero, then we have $Q_{\ell}(\DistOptUpdatedState^{\ast}_{\ell})=0$ for all $\ell\in\mathcal{R}$. As a result,
\begin{equation}\label{eq::convergence::coupled}
    \norm{\DistOptUpdatedState^{+}- \DistOptUpdatedState^{\ast}} \leq  O(\norm{\hat{\DistOptLocalState} - \DistOptUpdatedState^{\ast}}^2)
\end{equation}

The statement of \textit{Theorem~\ref{thm::convergence}} follows by combining of\eqref{eq::convergence::inexat::nlp}, \eqref{eq::convergence::decoupled} and \eqref{eq::convergence::coupled}.


\section{Open-source Implementation}\label{sec::implementation}

Based on the \autoref{alg:gn-aladin}, we improve the existing toolkit \acrshort{rapidpf}. To this end, in this section, we introduce a reduced modelling method and describe the structural upgrade of \acrshort{rapidpf+} compared with \acrshort{rapidpf}.

\subsection{Reduced modelling method}

\autoref{tb:bus type} summarizes the known and unknown variables of a \ac \acrshort{pf} problem according to different bus-types in the power system. In the original distributed \ac \acrshort{pf} model proposed by~\cite{muhlpfordt2021distributed}, the known variables are constrained by \textit{bus specification}, which is added as residuals in least-squares formulation. This results in the unnecessary growth of the problem dimension and slows down the run time. To overcome the issue, the present paper distinguishes the known and the unknown variables, and uses a so-called reduced modelling method to reduce the dimension of the distributed \ac \acrshort{pf} problem.
\begin{table}[htbp!]
    \begin{center}
    \caption{Known and Unknown variables for \ac \acrshort{pf} problem regarding the bus-type}\label{tb:bus type}
        \begin{tabular}{cccc}
                          & \slack           & \pq              & \pv \\\hline
        Known variables   & $\vang$, $\vmag$ & $\p$, $\q$       & $\vmag$, $\p$ \\
        Unknown variables & $\p$, $\q$       & $\vang$, $\vmag$ & $\vang$, $\q$ \\ \hline
        \end{tabular}
    \end{center}
\end{table}

For a specific region $\ell \in \mathcal{N^\textrm{reg}}$, the state consists of variables from both core buses and copy buses. The state of the core bus $i$ is defined according to its own bus-type:
\begin{align}
        \zeta^\textrm{core}_i=\begin{cases}
        (\p_i^\textrm{core}, \q_i^\textrm{core})  &\text{(\slack)}\\
        (\vang_i^\textrm{core}, \vmag_i^\textrm{core})  &\text{(\pq)}\\
        (\vang_i^\textrm{core}, \q_i^\textrm{core})  &\text{(\pv)}\\
    \end{cases},\; \forall i \in \setBus^\textrm{core}_{\ell},
\end{align}
whereas the state of the \textit{copy bus} $j$ contains voltage angle and magnitude  
\begin{equation}
    \zeta^\textrm{copy}_j = (\vang_j^\textrm{copy}, \vmag_j^\textrm{copy}),\;\forall j \in \setBus^\textrm{copy}_{\ell},
\end{equation}

The state of this specific region $\chi_{\ell}\in\mathbb{R}^{2\ncore_{\ell}+2\ncopy_{\ell}}$ is composed by all the core and the copy buses in the regions.

Typically, $\ncore$ dominates $\ncopy$ in a sub-system of a power grid. Therefore, the dimension by using the reduced modelling method, i.e., $\sum_{\ell} 2\ncore_{\ell}+2\ncopy_{\ell}$, is almost reduced by half, compared with the original model--- $\sum_{\ell} 4\ncore_{\ell}+2\ncopy_{\ell}$---proposed by~\cite{muhlpfordt2021distributed}. 

\subsection{rapidPF vs. rapidPF+}

\begin{figure}[htbp!]
    \centering
    \includegraphics[width=0.3\textwidth]{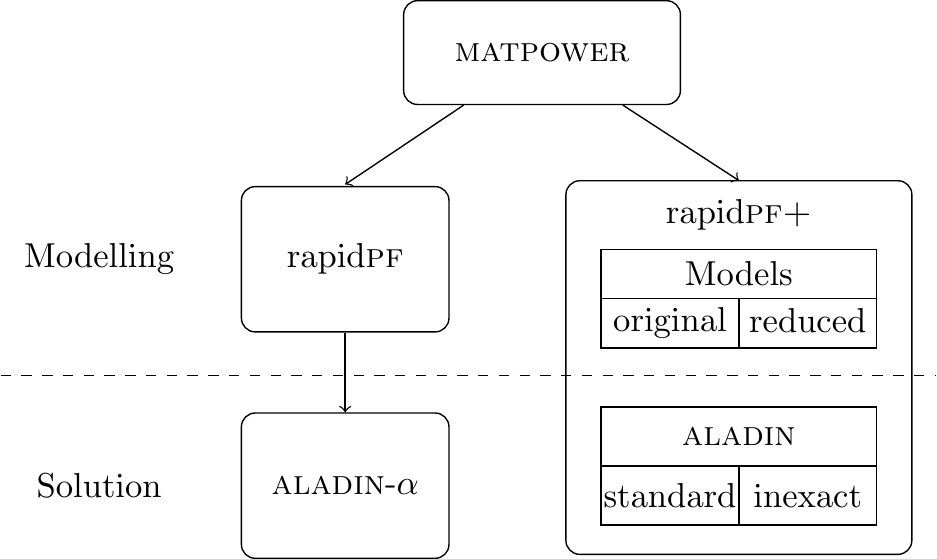}
    \caption{Flow charts for solving distributed \ac \acrshort{pf} by the \acrshort{rapidpf} and the \acrshort{rapidpf+} toolbox}\label{fig:flow chart}
\end{figure}

As shown in \autoref{fig:flow chart}, the \acrshort{rapidpf} builds a distributed \ac \acrshort{pf} problem based on \matpower case files and solves it by interfacing with an external \aladinalpha toolbox. Nevertheless, due to the problematic code efficiency of the \aladinalpha toolbox, computing for a large-scale problem is not acceptable---for a 4662-Bus system, it takes 90.1 seconds to converge by using \fmincon, whereas the initial time by using \casadi is intolerant.

In contrast, \acrshort{rapidpf+} doesn't rely on the external \acrshort{aladin} toolbox. The user can switch between two models and two \acrshort{aladin} algorithms. Comparison of these combinations is carried out in the following section.

\section{Simulation Results}\label{sec:results}

In this section, we illustrate the performance of several combinations of the two distributed \ac \acrshort{pf} models and the two variants of \acrshort{aladin} algorithm. We use the suggested combination by \cite{muhlpfordt2021distributed} as a benchmark, i.e., the original distributed power flow model with standard \acrshort{aladin}~(\autoref{alg:standard-aladin}). Towards practical implementation, several test cases by \cite{muhlpfordt2021distributed} are also modified---multiple connecting tie lines are added and the graph of regions is transferred from radial to meshed topology. Besides, we introduce a 10224-bus test case to exhibit the performance for a large-scale implementation.

The framework\footnote{The code is available on https://github.com/xinliang-dai/rapidPF} is built on \matlab-R2021a and the case studies are carried out on a standard desktop computer with \texttt{Intel\textsuperscript{\textregistered} 
i5-6600K CPU @ 3.50GHz} and 16.0 \textsc{GB} installed \textsc{ram}. The \casadi toolbox~\citep{andersson2019casadi} is used in \matlab, and \ipopt~\citep{wachter2006implementation} is used as the solver for decoupled \acrshort{nlp}s. To solve the linear system, a conjugate-gradient technique~\cite[Algorithm  7.2]{nocedal2006numerical} is implemented in order to avoid matrix-matrix multiplications, i.e., $J^{\top} J$.

Following~\cite{engelmann2018toward}, the quantities in the following are used to illustrate the convergence behavior
\begin{enumerate}
\item The deviation of optimization variables from the optimal value $\norm{x-x^*}_\infty$.
\item The primal residual, i.e., the violation of consensus constraint $\norm{A x -b}_\infty = \norm{\sum_{\ell\in\mathcal{R}}A_\ell x_\ell-b}_\infty$.
\item The dual residual $\gamma = \max_{\ell\in\mathcal{R}}\norm{\Sigma_{\ell}(\DistOptLocalState_\ell -\DistOptUpdatedState_\ell )}_\infty$. 
\item The solution gap calculated as $\left|f(x)-f(x^*)\right|$, where $f(x^*)$ is provided by the centralized approach.
\end{enumerate}

\subsection{Comparison of different combinations}

For fair comparison, the primal variables $x$ are initialized with the initial guess provided by \matpower \citep{zimmerman2010matpower}, while the dual variable $\lambda$ is set to zero. The tuning parameters $\rho$ and $\mu$ of the \acrshort{aladin} algorithm are set to $10^{2}$, whereas the tolerance $\epsilon$ is set to $10^{-8}$. \runpf from \matpower is used to represent a centralized approach.

\autoref{tb:results} displays the computing time of different combinations. The computing time of both algorithms also benefit from the dimensional reduction---compared with the original distributed \ac \acrshort{pf} model, the dimension by applying reduced modelling method is decreased almost by half.

What else stands out in this table is the fast computing time of the Gauss-Newton based inexact \acrshort{aladin}~(\autoref{alg:gn-aladin}). In contrast to solving \acrshort{nlp} in a decoupled step of \autoref{alg:standard-aladin}, \autoref{alg:gn-aladin} solves the equivalent linear systems of a quadratic approximation in both decoupled and coupled steps by exploiting the structure of the problem formulation. Consequently, the computation effort has been reduced dramatically. As a result, the computing time of solving the reduced distributed \acrshort{pf} model by using \autoref{alg:gn-aladin} is in the same order of magnitude with the centralized approach, and can be further improved by implementing parallel computing.

\subsection{Convergence behavior of 10224-Bus system}

\begin{figure}[htbp!]
    \centering
    \includegraphics[width=0.15\textwidth]{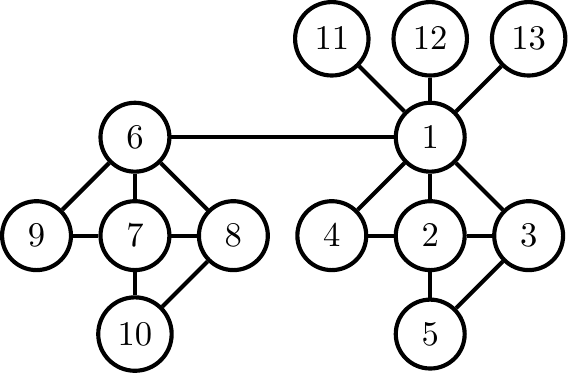}
    \caption{Connection graph of 10224-bus test case}\label{fig:10224} 
\end{figure}

Next, we study the convergence behavior of the largest test case, i.e., 10224-bus system. The test case is composed of six 1354-bus \matpower test cases, and seven 300-bus \matpower test cases. Its connection graph of regions are shown in~\autoref{fig:10224}. 

\begin{figure*}[htbp!]
    \centering
    \includegraphics[width=0.7\textwidth]{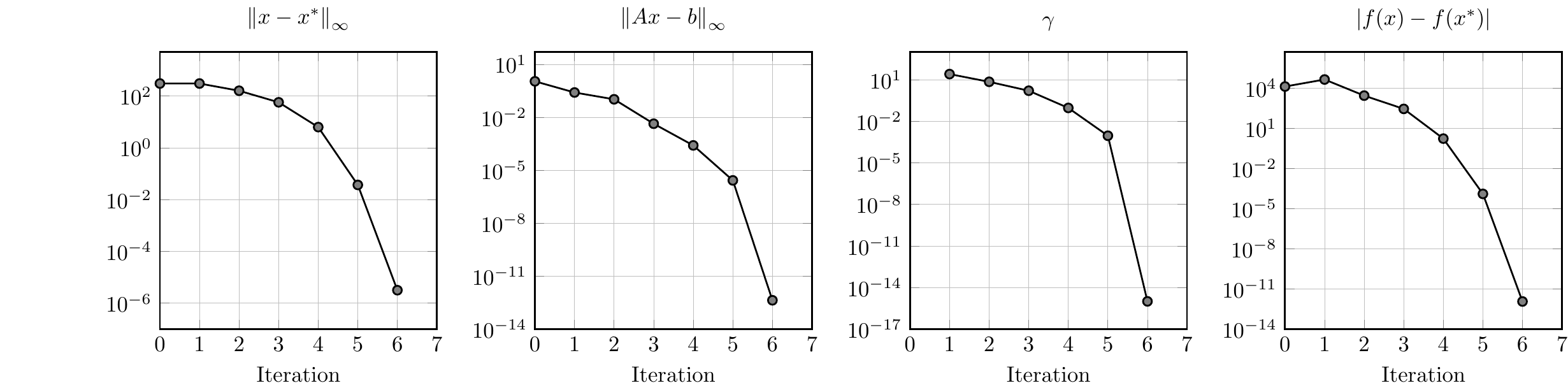}
    \caption{Convergence behavior of 10224-bus system by applying reduced modeling method with the Gauss-Newton based inexact \acrshort{aladin} algorithm}\label{fig::convergence} 
\end{figure*}

To solve the \ac \acrshort{pf} problem of the 10224-Bus system, we use the reduced modelling method with the Gauss-Newton based inexact \acrshort{aladin} algorithm. \autoref{fig::convergence} shows the four quantities in every iterate, i.e., the deviation of current variables from the optimal value, the primal residual, the dual residual and the solution gap. Within half a dozen iterates, the new \acrshort{aladin} algorithm converges to the optimal solution with high accuracy, as presented in~\autoref{tb:deviation}. At the same time, a locally quadratic convergence rate can be observed from \autoref{fig::convergence}.

\begin{table}[htbp!]
    \renewcommand\arraystretch{1.5}
    \caption{The deviation of the solution for 10224-bus system from the optimal value by applying reduced modeling method with the Gauss-Newton based inexact ALADIN}\label{tb:deviation}
    \centering
    \begin{tabular}{rrrrr}\linespread{1.5} 
        & \vang [rad] & \vmag [p.u.] & \p[p.u.] &\q[p.u.]\\\hline
        $\norm{\;\cdot\;}_\infty$  & $1.7\times10^{-8}$ & $7.5\times10^{-9}$ & $5.7\times10^{-7}$ & $3.2\times10^{-6}$\\\hline
    \end{tabular}
\end{table}

\begin{table*}[hbtp!]
    \centering
    \caption{Computing time for solving power flow problem with different combinations}\label{tb:results}
  \begin{tabular}{rrrrrrrrrrr}
    & & &  \multicolumn{3}{c}{Original model} & \multicolumn{3}{c}{Reduced model} &  \\
    \multirow{-2}{*}{Buses} & \multirow{-2}{*}{\nregions} & \multirow{-2}{*}{\nconnections} & Dimension &  standard[s] & inexact[s] &  Dimension &  standard[s] & inexact[s] &  \multirow{-2}{*}{centralized}\\\hline
    53    &   3 & 5   &   232   & 0.143 & 0.027 & 126   & 0.114 &    0.011 & 0.004\\ 
    418   &   2 & 8   &   1704  & 0.485 & 0.068 & 868  & 0.315 &    0.028 & 0.014\\
    2708  &   2 & 30  &   10952 & 3.913 & 0.236 & 5536 & 2.149 &   0.109 & 0.051\\ 
    4662  &   5 & 130 &   19168 & 10.442& 0.451 & 9844 & 5.694 &   0.228 & 0.129\\
    10224 &  13 & 242 &   41864 & 25.909      & 0.996 & 21416&    14.392   &   0.591 & 0.257 \\\hline
  \end{tabular}
\end{table*}

\section{Conclusions}

The present paper investigates the application of a new tailored version of \acrshort{aladin} for solving the \ac \acrfull{pf} problem. Compared with the previous work by~\cite{muhlpfordt2021distributed}, the dimension by applying reduced modelling method can be reduced by half. By applying the Gauss-Newton based inexact \acrshort{aladin}, we trade off the convergence rate slightly for the great improvement on computing time. Besides, no external \acrshort{nlp} solver is needed. In general, this new combination is of great potential for handling large-scale systems, and turns out to be as efficient as a centralized approach. For future work, efforts toward parallel computing will be made to reduce the computing time even further.

\bibliographystyle{IEEEtranN}
\bibliography{main}

\end{document}